\documentclass[manuscript]{aastex}
\usepackage{lscape}

\shorttitle{The spectral energy distribution of the Carina nebula
from Far Infrared to Radio wavelengths } \shortauthors{Salatino et
al.}

\begin{document}

\title{The spectral energy distribution of the Carina nebula \\ from far infrared to radio wavelengths}

\author{M. Salatino, P. de Bernardis, S. Masi}
\affil{Physics Department, Sapienza Universit\`{a} di Roma, p.le Aldo Moro 2, 00185 Roma, Italy}
\email{maria.salatino@roma1.infn.it}

\and

\author{G. Polenta}

\affil{ASI Science Data Center, ESRIN, via G. Galilei, 00044, Frascati, Italy}
\affil{INAF-Osservatorio Astronomico di Roma, via di Frascati 33, I-00040 Monte
Porzio Catone, Italy}

\begin{abstract}
Multi-wavelength observations are mandatory to understand the physical properties of astrophysical sources. In this paper we use
observations in the far infrared to radio range to derive the spectral energy distribution (SED) of the Carina nebula. To do this,
we carefully subtract the irregularly varying diffuse emission from the Galactic plane, which can be of the order of 10$\,\%$ of
the nebula flux at these wavelengths. We find that the far infrared SED can be modeled as emission from a dust population with a
single temperature $T_d=(34.5^{+2.0}_{-1.8})$, and with a spectral index of emissivity $\alpha=-1.37^{+0.09}_{-0.08}$. We also find
a total infrared luminosity of the Nebula of $(7.4^{+2.5}_{-1.4})\times10^6 L_{\odot}$ and, assuming a single temperature of the
dust, a mass of the dust of $(9500^{+4600}_{-3500})M_\odot$.
\end{abstract}


\section{Introduction}\label{sec intro}
The Carina nebula (NGC 3372 or RCW 53) has one of the highest surface brightnesses of any nebula in the southern hemisphere. It
contains the most luminous and one of the most massive stars of our Galaxy: eta Carinae, a LBV star. Its distance (2.3$\,$kpc) is
known accurately, mainly through the expansion parallax \citep{Allen93} of the Homunculus ne\-bu\-la \citep{Smith05a}. The nebula
has a particularly low extinction with respect to other star formation regions, and presents an anomalous reddening law
\citep{Smith87}. The nebula hosts, moreover, more than 60 O stars, three WNL stars, an O2 supergiant, and a B1.5 supergiant which
suggests the presence of a proto OB association \citep{Smith06}. The Carina nebula offers a unique laboratory to study different
astrophysical problems: star formation, feedback on the star formation provided by young massive stars, as well as triggering of a
second generation of star formation from the most massive ones; small-scale phenomena (evaporating protoplanetary disks, jets,
cometary clouds, dust pillars, globules, Herbig-Haro jets \citep{Smith04}, and photodissociation regions on the surfaces of
molecular clouds \citep{Rathborne02}). High-quality data spanning from infrared (IR) to radio wavelengths show a heavily active
star formation region in which the LBV star, as well as other massive ones, play a crucial role.

Knowledge of the nebula improves our knowledge of the interstellar medium, and, in particular, our models of interstellar dust.
Observational cosmology has an interest in understanding dust properties too \citep{Deschenes08,Dunkley08,Lazarian03,Masi01}:
current measurements of the po\-la\-ri\-za\-tion of the cosmic microwave background (CMB) are limited by our limited ability to
subtract foreground emission. Future experiments like the \emph{Planck} and \emph{Herschel} satellites (launched on 2009 May 14),
and the PILOT \citep{Bernard07a,Bernard07b}, EBEX \citep{Oxley04,Reichborn10}, SPIDER \citep{Montroy06}, and Balloon Observations
of Millimetric Extragalactic Radiation and Geophysics (BOOMERanG) balloons, with their polarization surveys of the sky will provide
data to understand dust properties: like the emissivities and the wavelength-dependence of the far infrared (FIR) polarized
emission \citep{Hildebrand99,Lazarian07,Martin07}. This knowledge is very much needed for a better extraction of the finest CMB
polarization properties, which can provide information on cosmological inflation, in post-\emph{Planck} missions like CMB-Pol
\citep{Baumann09}, B-Pol \citep{deBernardis09} and COrE (http:$//$www.core-mission.org).

The paper is organized as follows: in Section$\,$\ref{sec Smith analysis}, we briefly summarize the current knowledge of the Carina
nebula. In Section$\,$\ref{sec multi l analysis} we present the data we used for our analysis. The subtraction of the background is
shown to be, with the calibration error, the dominant source of error for the measured fluxes (Section$\,$\ref{sec bckg}). In
Section$\,$\ref{sec SED}, we present the FIR to radio spectral energy distribution (SED), and the best fit we performed using
$\chi^2$ minimization and a Monte Carlo Markov Chain Metropolis-Hastings (MCMCMH) algorithm for the estimate of the confidence
intervals. We derive some global properties of the nebula and compare our results to previous literature data in
Section$\,$\ref{sec discus}, and we conclude summarizing the results and future activities in Section$\,$\ref{sec concl}.

\section{The SED and the global properties}\label{sec Smith analysis}

\citet{Smith07} have integrated \emph{IRAS} (FIR) data and the Parkes (radio) data to estimate the flux of the Carina nebula in a
box $1^{\circ}58^{\prime}\times2^{\circ}45^{\prime}$ wide. The resulting SED is described by three graybody components having
temperatures of 220, 80, and 35$\,$K with emissivity $\propto \lambda^{-1}$, and by an optically thin thermal bremsstrahlung radio
continuum emission component ($\propto\nu^{-0.1}$). Their analysis used the two data at 60 and 100$\,\mu$m to fit the Carina nebula
SED in the wavelength range (20-5000)$\,\mu$m, with two free parameters (temperature and emissivity), assuming a spectral index of
emissivity -1; the single measurement at radio wavelengths is used to measure the amplitude of the thermal bremsstrahlung
component.

Luminosity at these wavelengths is independent of dust grains radius and emissivity as long as the radius of the grains is small,
$< 0.2 \mu$m \citep{Draine84}. With a typical grain density, $\rho$, of about 3g$\,$cm$^{-3}$ and a temperature $T$, the dust mass
which emits an IR luminosity, $L_{IR}$, is \citep{Smith05}:
\begin{equation}\label{m dust}
    M_{dust}(M_{\odot})=\frac{122\rho}{3 \sigma_{SB} T^6} L_{IR}(L_{\odot}),
\end{equation}
with $\sigma_{SB}$ being the Stefan-Boltzmann constant expressed in cgs units. For the dust population at 35$\,$K \citet{Smith07}
found a luminosity and a mass equal to 7.6$\times10^6\,L_{\odot}$ and 9600 $M_{\odot}$, respectively. In the following we keep the
same area considered by \citet{Smith07} and we estimate the integrated flux of additional experiments carried out at different
wavelengths.

Studies of the Carina nebula at wavelengths shorter than 60$\mu$m are already present in literature
\citep{Cox95,Morris99,Rathborne02,Smith07}. In this work we demonstrate that in the FIR portion of the spectrum the B98 and
\emph{Wilkinson Microwave Anisotropy Probe} (\emph{WMAP}) data improve our knowledge of the nebula.

\section{Multiwavelength observations}\label{sec multi l analysis}

In this section, we describe the multiwavelength data we used and how they have been analyzed. Although we used different
experiments, some parts in the data analysis are the same.

For each experiment the average value of the background has been estimated as described in Section$\,$\ref{sec bckg}. For the
background estimation, we have also considered a possible contribution from CMB anisotropy, but we found that it is negligible,
given the large number of independent sky regions present in our integration box.

\subsection{COBE}
The DIRBE instrument of the COBE satellite has produced infrared absolute sky brightness maps of the Carina nebula in the
wavelength range 1.25 to 240$\,\mu$m with an instantaneous field of view of 0.$^{\circ}$7$\times$0.$^{\circ}$7 \citep{Hauser98a}.
For our analysis we used data at wavelengths longer than 60$\mu$m. The quadrilateralized spherical cube projection of the DIRBE
maps has been converted into the Hierarchical Equal Area isoLatitude PIXelization, HEALPIX \citep{Gorski05}; this has required a
change of pixel size with respect to the original: from about 0.$^{\circ}$32o to $27.^{\prime}5$ (nside=128). From these data we
have calculated the maps and the integrated fluxes in the same box considered by \citet{Smith07}. The results are shown in
Figure$\,$\ref{fig1} and Table$\,$\ref{Tab fluxes ncase5}. The errors in the calculated fluxes are the quadrature sum of the
uncertainty in the absolute gain ca\-li\-bra\-tion, the uncertainty in the instrumental offset (zero point), the standard deviation
of the weighted average photometry values \citep{Hauser98b}, and the contribution from both the background diffuse emission of the
Galaxy (see Section$\,$\ref{sec bckg}) and the rms CMB anisotropy.

\subsection{IRAS}
IRAS has observed, at wavelengths of 60 and 100 $\mu$m, the Carina nebula with an angular resolution of
1.$^{\prime}5\times4.^{\prime}7$, and 3.$^{\prime}0\times$5.$^{\prime}0$, respectively \citep{Neugebauer84}. We have used the
latest processed IRAS images IRIS \citep{Deschenes05}: they have an absolute calibration coherent with DIRBE, a better zodiacal
light subtraction, and better destriping with respect to the original ISSA maps \citep{Wheelock93}. A preliminary comparison with
the original ISSA data gave flux values 22$\%$ and 7.9$\%$ smaller than the IRIS data at 60 and 100$\,\mu$m, respectively, the
discrepancy with the IRIS data is 1.5$\,$(0.3)$\,\sigma$ at 60 (100)$\,\mu$m. For coherence with the other experiments, that are in
the HEALPIX pixelization, we used the IRIS data. The resolution is 1.$^{\prime}7$ (nside 2048) \citep{Gorski05}. The brightness map
in our integration box is reported in Figure$\,$\ref{fig1} and the resulting flux in Table$\,$\ref{Tab fluxes ncase5}.

\subsection{BOOMERanG 98}

\begin{figure}[h]
\epsscale{0.8}\plotone{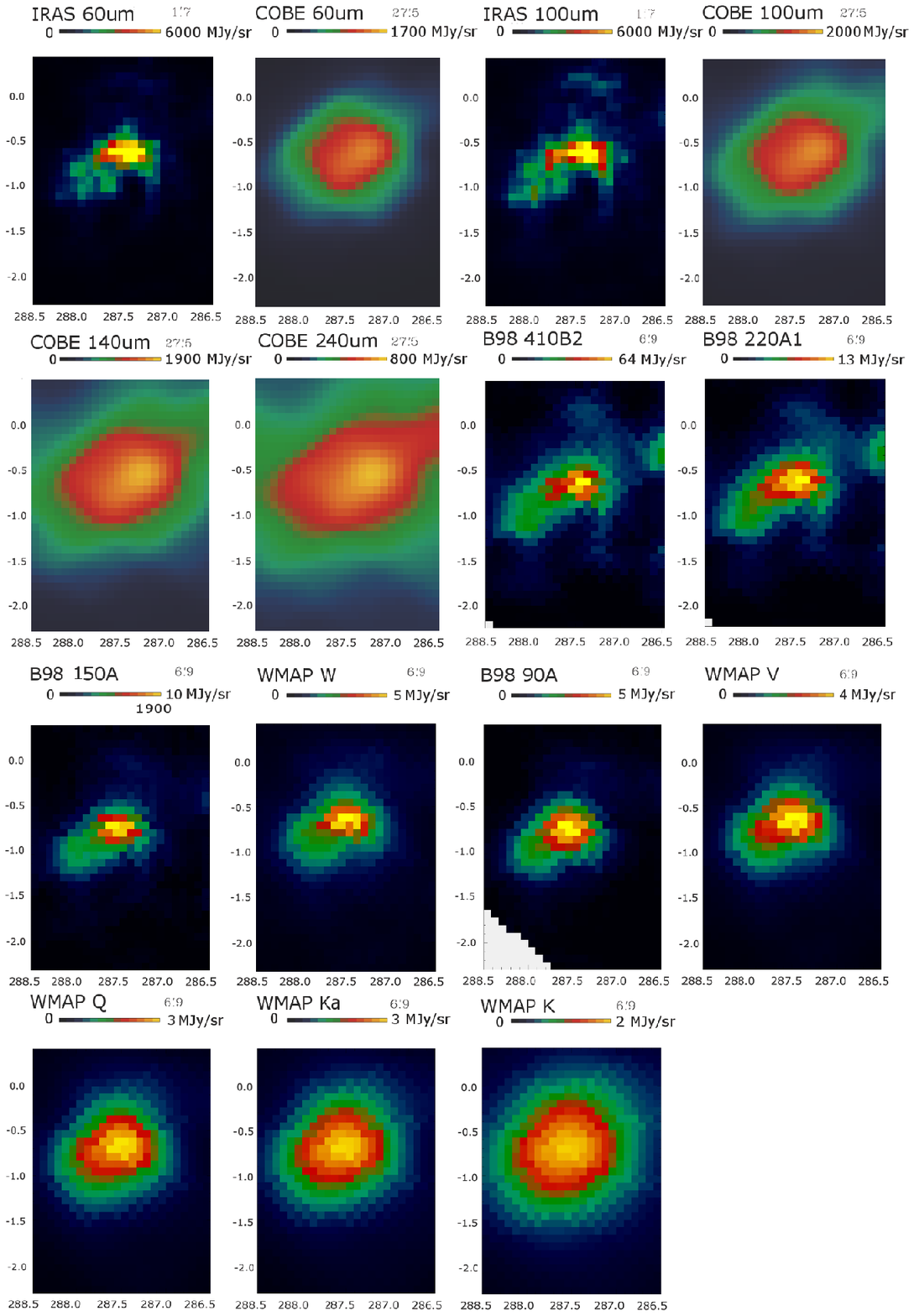}\caption{Wide-field images of the Carina nebula, from far-infrared (top left) to radio (bottom right)
wavelengths. The ranges of the galactic coordinates (in degrees) reproduce the \citet{Smith07} integration box.} \label{fig1}
\end{figure}

The stratospheric balloon experiment BOOMERanG 98, hereafter B98, \citep{Crill03} observed the Carina nebula at frequencies of 90,
150, 220, and 410$\,$GHz \citep{Coble05} with an angular resolution of about 18.$^{\prime}1$, 11.$^{\prime}7$, 15.$^{\prime}9$ and
15.$^{\prime}0$, respectively. For the 410$\,$GHz channel the calibration error is 25$\%$ \citep{Masi01}, for the others 10$\%$
\citep{Ruhl03}. Since the B98 survey does not fill completely the Smith integration box, the missing region (white region in B98
maps, Figure$\,$\ref{fig1}) is filled with the average signal from an observed background region of area
0.$^{\circ}2\times$0.$^{\circ}2$ (see Tab.$\,$\ref{Tab B98 boxes}). The missing area is at most 10$\%$ of the integration box, in a
low background region. From the WMAP7 map at a similar frequency (94$\,$GHz) we see that neglecting completely the contribution
from this area the flux decreases by 1.2$\%$, so the error we introduce with our procedure is certainly less than 1$\%$ and in any
case is smaller than the errors introduced by the background subtraction procedure (see Section$\,$\ref{sec bckg}). Fluxes have
been corrected for the instrumental response of the telescope using the transmissions functions of the 16 B98 filters. The signal,
in Jansky, for each channel $i$, $S_{i}$, is given by
\begin{equation}\label{B98 maps formula}
    S_{i}=\frac{k_{A,i} T_{CMB}}{\int_{\Delta\nu_i} e_i(\nu)B(\nu)\frac{x e^x}{e^x-1} \Delta \Omega d\nu}\int_{\Delta\nu_i} e_i(\nu)I_i(\nu)d\nu,
\end{equation}
in which $T_{CMB} = 2.73\,$K is the temperature of the CMB \citep{Fixsen96}, $x=h\nu/(kT)$ (with $k$ being the Boltzmann constant
and $h$ being the Planck constant), $k_{A,i}$ is the calibration factor of the $i$th channel, $e_i(\nu)$ the filter transmission
function of the $i$th channel, and $I_i(\nu)$ is the theoretical emission model.

\begin{table}[h!]
\begin{center}
\caption{Regions Observed by B98 in Different Frequency Channels, and Used to Fill Missing Regions in the B98 Survey; the Columns
List Respectively the B98 Channels, and the Galactic Coordinates of the Edges of the Regions: $\ell_{min}$, $\ell_{max}$,
$b_{min}$, $b_{max}$.} \label{Tab B98 boxes}
\begin{tabular}{ccccc}
\tableline\tableline
Channel                              & $\ell_{min}$         & $\ell_{max}$     & $b_{min}$      & $b_{max}$      \\
\tableline
B90A, B90B                           &  287$^{\circ}$.9     &  288$^{\circ}$.1 & -1$^{\circ}$.8 & -1$^{\circ}$.6 \\
B150A, B150B, B150A2, B150B1, B150B2 &  288$^{\circ}$.3     &  288$^{\circ}$.5 & -2$^{\circ}$.1 & -1$^{\circ}$.9 \\
B150A1                               &  287$^{\circ}$.6     &  287$^{\circ}$.8 & -2$^{\circ}$.1 & -1$^{\circ}$.9 \\
B220A1, B220B1, B220B2               &  287$^{\circ}$.9     &  288$^{\circ}$.1 & -1$^{\circ}$.8 & -1$^{\circ}$.6 \\
B220A2                               &  288$^{\circ}$.3     &  288$^{\circ}$.5 & -2$^{\circ}$.1 & -1$^{\circ}$.9 \\
B410A2, B410B1                       &  288$^{\circ}$.3     &  288$^{\circ}$.5 & -2$^{\circ}$.1 & -1$^{\circ}$.9 \\
B410A1                               &  288$^{\circ}$.3     &  288$^{\circ}$.5 & -1$^{\circ}$.8 & -1$^{\circ}$.6 \\
\hline
\end{tabular}
\end{center}
\end{table}

We have assumed a typical power law, $I_i(\nu)=I_{0,i}(\nu / \nu_{0,i})^{\beta}$, where $I_{0,ith}$ is the flux measured at
frequency $\nu_{0,i}$, the central frequency of the $i$th channel. The integration is performed over the entire band of the filter,
$\Delta\nu_i$. The color corrections (Table$\,$\ref{Tab B98 fluxes}) are less than 1$\%$ for an optically thin thermal
bremsstrahlung ($\beta=-0.1$), Rayleigh-Jeans spectrum ($\beta=2$) or a power law with $\beta=3$. Since in the B98 wavelength range
(excluding the 90$\,$GHz channels) we expect thermal dust emission to be predominant over bremsstrahlung emission, we carried our
analysis with the color corrections relative to $\beta=2$. B98 was a scanning instrument. To remove instrumental offsets and
drifts, we have high-pass filtered the time-ordered data. The cutoff of the high-pass filter corresponds to an angle $\theta_C$ in
the sky, so the B98 maps we have used do not contain structures at scales larger than $\theta_C$. We have verified that the fluxes
in our target region are not significantly affected by filtering if $\theta_C \gtrsim 10^{\circ}$ (see Table$\,$\ref{Tab B98
fluxes}).

\begin{table}
\begin{center}
\caption{Variations of the Integrated Fluxes (Color-corrected) Measured at 150$\,$GHz for Different Spectral Index of SED, $\beta$,
and Different Cutoff $\theta_C$ of the High-pass Filter Applied to the Time-ordered Data. The fluxes are corrected for the
background emission ($n=5$, see Section$\,$\ref{sec bckg}).} \label{Tab B98 fluxes}
\begin{tabular}{ccc}
\tableline\tableline
Filter         & $\beta$          &  Flux (Jy)      \\
\tableline
$5^{\circ}.0$  & -0.1             &  $920  \pm 150$ \\
               & 2.0              &  $920  \pm 150$ \\
               & 3.0              &  $910  \pm 150$ \\
$10^{\circ}.0$ & -0.1             &  $1630 \pm 190$ \\
               & 2.0              &  $1620 \pm 190$ \\
               & 3.0              &  $1610 \pm 190$ \\
$15^{\circ}.0$ & -0.1             &  $1920 \pm 210$ \\
               & 2.0              &  $1920 \pm 210$ \\
               & 3.0              &  $1900 \pm 210$ \\
\tableline
\end{tabular}
\end{center}
\end{table}

The errors in the fluxes are the quadrature sum of the standard deviations of the measured signals, the calibration uncertainties,
and the Galactic background contributions. Depending on the channel, the calibration error dominates on the background one and vice
versa (Table$\,$\ref{Tab fluxes ncase5}) while the statistical error represents a small contribution to the total.

The final integrated fluxes are computed as averages of different channels with the same frequency: one for 90$\,$GHz, four for
150$\,$GHz, three for 220$\,$GHz, and three for 410$\,$GHz. The results are reported in Table$\,$\ref{Tab fluxes ncase5} and
Figure$\,$\ref{fig1}.

\subsection{WMAP}
The Carina nebula has been observed by the WMAP satellite \citep{Bennett03} at the frequencies: 23$\,$GHz (\emph{K}-band),
33$\,$GHz (\emph{Ka}-band), 41$\,$GHz (\emph{Q}-band), 61$\,$GHz (\emph{V}-band) and 94$\,$GHz (\emph{W}-band); the seven-year
release data, hereafter WMAP7, provide maps of the nebula with FWHM angular resolution of about $0^{\circ}.81$, $0^{\circ}.62$,
$0^{\circ}.48$, $0.^{\circ}33$, and $0.^{\circ}21$, respectively \citep{Hinshaw07,Hinshaw09}. We used the temperature (Stokes $I$
parameter) maps per frequency band obtained from the Lambda archive (http://lambda.gsfc.nasa.gov/). The fluxes have been corrected
for the instrumental responses of each of the 20 differential radiometers; we averaged the bandpass frequency responses of all the
radiometers in each frequency band. The errors have been estimated as the quadrature sum of statistical error, background
uncertainty, calibration error ($0.5\%$) and pixel noise \citep{Hinshaw09}. The results for the integrated fluxes are reported in
Table$\,$\ref{Tab fluxes ncase5} and the maps in Figure$\,$\ref{fig1}.

\subsection{Spitzer}
The MIPS instrument on board of the \emph{Spitzer} satellite (Spitzer archive: http://archive.spitzer.caltech.edu/) has observed
the Carina nebula at far infrared wavelengths of 24, 70, and 160$\,\mu$m \citep{Rieke04}. However, the photoconductors of MIPS were
significantly saturated by the strong emission of the nebula, so we decided to exclude these data from our analysis.

\subsection{Multiwavelength maps}

The maps resulting from our analysis of all the experiments are displayed in Figure$\,$\ref{fig1}; they are labeled with their
HEALPIX pixel scale. The COBE maps, nside=128, have been filtered with a Gaussian low-pass filter (FWHM=0$^{\circ}$.35). Because of
the low resolution of the instruments, the inner morphology of the Carina nebula is not resolved in the COBE-DIRBE maps and in the
three lower frequency channels of WMAP. The lower resolution of the B98 instrument, with respect to the experiments analyzed by
\citet{Smith07}, does not allow a real pixel to pixel comparison: there is, nevertheless, significant morphological resemblance
between the IRAS-IRIS, the B98 maps and the 61 and 94$\,$GHz channels of WMAP7.

\section{Background subtraction}\label{sec bckg}

The region around the Carina nebula, near the Sagittarius-Carina spiral arm, is rich in massive star formation and HII regions: NGC
3603, NGC 3576, NGC 3572, NGC 3324, NGC 3293 and RCW 49 \citep{Smith08}. With respect to these regions Carina is larger, brighter,
closer (2.3 kpc against 6-8 kpc) and has a lower extinction and reddening; therefore, its emission should dominate.

Background contributions (CMB, sky and  irregularly varying diffuse emission) have been studied averaging flux values with two
different numbers of background regions (labeled with the index \emph{n}). In both cases we use six sets of reference regions, each
set including five regions ($n=5$) and two regions ($n=2$). All regions have an area of $0.^{\circ}5\times0.^{\circ}5$ (see
Table$\,$\ref{Tab bckg coord}). For a given $n$, the reference regions are labeled with the index \emph{m}. For $n=5$ the regions
have been randomly chosen around the Carina nebula, for $n=2$ the regions are close to the nebula and symmetrically located with
respect to its center. Only regions observed by all the experiments and close to the Carina nebula have been selected. We computed
the average of the fluxes of all the regions in each set. We used the average of all the sets as an estimate of the background, and
its standard deviation as an estimate of the error on the background.

\begin{table}[h!]
\begin{center}
\caption{Selected Sky Areas for the Study of the Diffuse Background Contribution. Top part: \emph{n=5}, bottom part: \emph{n=2};
\emph{m} is the index of the region, $\ell_{i}$, $b_i$ ($i$=1,5) are the galactic coordinates (in degrees) for the centers of the
regions.}  \label{Tab bckg coord}
\begin{tabular}{ccccccccccc}
\tableline\tableline
$m$ &$\ell_1$&  $b_1$& $\ell_2$& $b_2$& $\ell_3$& $b_3$& $\ell_4$& $b_4$& $\ell_5$ & $b_5$\\
\tableline
1   & 289.75 &  1.25 & 285.15  & -3.75& 285.15  & 1.25 & 287.20  & 1.25 & 285.75   & -2.35\\
2   & 289.75 &  2.25 & 286.15  & -3.75& 285.15  & 2.25 & 287.20  & 2.25 & 285.75   & -3.35\\
3   & 289.75 &  3.25 & 284.75  & -2.25& 288.25  & 1.25 & 284.25  &-3.75 & 285.55   &  3.25\\
4   & 288.75 &  1.75 & 286.25  &  1.75& 284.25  &-4.25 & 285.25  &-2.25 & 287.75   &  3.75\\
5   & 288.25 &  3.25 & 284.25  & -2.75& 284.25  & 1.75 & 289.75  & 4.75 & 289.25   &  1.75\\
6   & 286.25 &  4.75 & 284.75  & -3.25& 284.25  & 1.25 & 290.25  & 2.75 & 284.75   &  2.95\\
\tableline
\end{tabular}
\begin{tabular}{ccccccccccc}
\tableline\tableline
$m$ &$\ell_1$& $b_1$& $\ell_2$& $b_2$\\
\tableline
1   & 285.25 & -3.75 & 289.75 & 2.25 \\
2   & 286.25 & -2.25 & 288.75 & 0.75 \\
3   & 284.75 & -3.25 & 290.75 & 1.75 \\
4   & 284.25 & -3.75 & 290.75 & 2.25 \\
5   & 285.75 & -3.25 & 289.25 & 1.75 \\
6   & 285.25 & -2.75 & 289.75 & 1.25 \\
\tableline
\end{tabular}
\end{center}
\end{table}

\begin{figure}[h!]
\epsscale{.80}\plotone{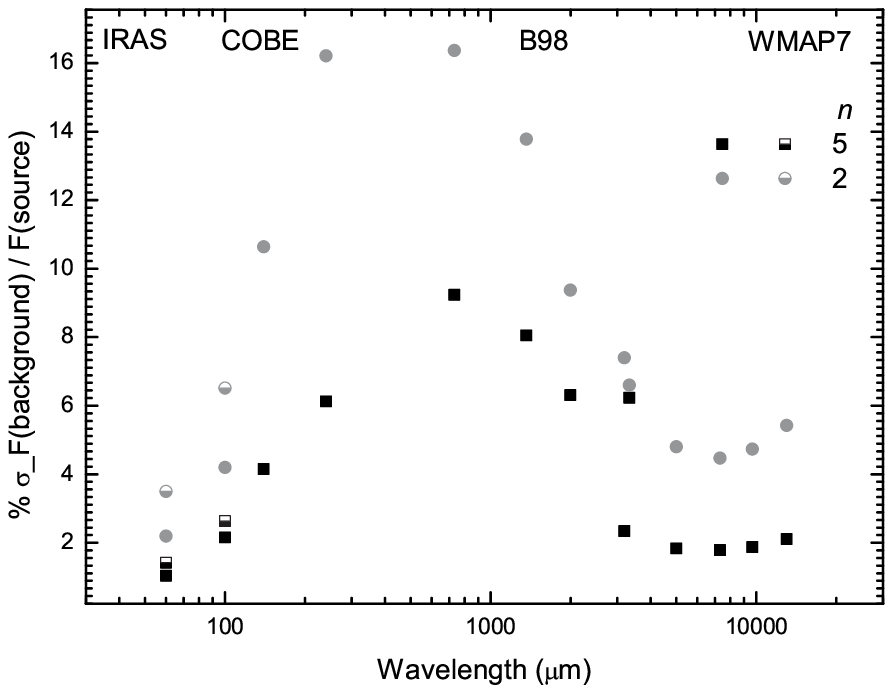} \caption{Standard deviation of the mean background contribution, $\sigma\_F(background)$, to the
measured flux, $F(source)$, corrected for the background, vs. the wavelength of the different data sets; the number (\emph{n})
labels the different number of regions considered for estimating the mean background. The \emph{IRAS} points are labeled with
half-colored symbols.} \label{fig2}
\end{figure}

An indicator of the effect of background fluctuations is the ratio between the standard deviation of the average background and the
nebula flux (corrected for the background). When computed for all the regions of Table$\,$\ref{Tab bckg coord} this indicator has a
trend increasing with wavelength, and is maximum between the 240$\,\mu$m DIRBE and the 410$\,$GHz B98 images; on the contrary this
indicator is basically independent of the considered sky region (Figure$\,$\ref{fig2}).

The background contribution is more dependent on the number of regions considered for the analysis; this seems to be true for all
the experiments. Similar background trends are found considering a small number of sky regions (maximum three) with different solid
angles (from $1^{\circ}\times1^{\circ}$ up to $6^{\circ}\times2^{\circ}$).

In both cases, the WMAP data are the least affected by the background, followed by B98, IRAS and COBE. The background
disagreement between the B98 90$\,$GHz and WMAP5 94$\,$GHz is due to the high-pass filtering of the BOOMERanG data.

Our analysis seems to indicate, for the irregularly varying emission surrounding the Carina nebula, a significative spectral
dependence and a large scale nature (larger than $6^{\circ}\times4^{\circ}$).

The error on the background in case $n=5$ is about 1.5 times smaller than in case n=2. Since case $n=5$ has better statistical
significance, we have selected this configuration for the subtraction of the background in our integrated flux analysis. However,
we have verified that all the results we give below remain very similar if we use the background estimates of case $n=2$.

\section{SED of the Carina Nebula, from FIR to cm-wavelengths}\label{sec SED}

\begin{table}
\begin{center}
\caption{Integrated Fluxes for \emph{n}=5. For a given wavelength of observation, $\lambda$, of an experiment, we report, in the
order: experiment, flux, total, statistical, background, and calibration errors.} \label{Tab fluxes ncase5}
\begin{tabular}{ccccccc}
\tableline\tableline
$\lambda$  &Experiment    &Flux (Jy)                &Error ($\%$)& Stat ($\%$) & Bckg ($\%$)& Cal ($\%$) \\
60  $\mu$m &IRAS (IRIS)   &$(7.5 \pm 0.8)\times10^5$&10.4    &0.04             & 1.0        & 10  \\
60  $\mu$m &COBE          &$(5.7\pm 0.6)\times10^5$ &10.5    &0.2              & 1.4        & 10  \\
100 $\mu$m &IRAS (IRIS)   &$(9.7\pm 1.3)\times10^5$ &13.7    &0.03             & 2.2        & 13  \\
100 $\mu$m &COBE          &$(8.0\pm 1.1)\times10^5$ &13.8    &0.2              & 2.6        & 13  \\
140 $\mu$m &COBE          &$(9.0\pm 1.0)\times10^5$ &11.4    &0.1              & 4.1        & 11  \\
240 $\mu$m &COBE          &$(4.1\pm 0.5)\times10^5$ &13.1    &0.1              & 6.1        & 12  \\
731 $\mu$m &B98           &$(1.6\pm 0.4)\times10^4$ &26.6    &0.2              & 9.2        & 25  \\
1.4 mm     &B98           &$(4.2\pm 0.5)\times10^3$ &12.8    &0.3              & 8.0        & 10  \\
2.0 mm     &B98           &$(1.6\pm 0.2)\times10^3$ &11.8    &0.6              & 6.3        & 10  \\
3.2 mm     &WMAP\textbf{7}&$(10.2\pm 0.2)\times10^2$&2.4     &1.1$\cdot10^{-3}$& 2.3        & 1.7 \\
3.3 mm     &B98           &$(0.9\pm 0.1)\times10^3$ &11.8    &0.5              & 6.2        & 10  \\
5.0 mm     &WMAP\textbf{7}&$(9.5\pm 0.2)\times10^2$ &1.9     &9.7$\cdot10^{-4}$& 1.8        & 1.6 \\
7.3 mm     &WMAP\textbf{7}&$(9.6\pm 0.2)\times10^2$ &1.9     &5.1$\cdot10^{-4}$& 1.8        & 1.6 \\
9.7 mm     &WMAP\textbf{7}&$(9.9\pm 0.2)\times10^2$ &1.9     &5.4$\cdot10^{-4}$& 1.9        & 1.6 \\
1.3 cm     &WMAP\textbf{7}&$(9.7\pm 0.2)\times10^2$ &2.0     &2.6$\cdot10^{-4}$& 2.1        & 1.6 \\
\tableline
\end{tabular}
\end{center}
\end{table}

The integrated fluxes, with the background subtraction described in Section$\,$\ref{sec bckg}, are reported in Table$\,$\ref{Tab
fluxes ncase5}. In Figure$\,$\ref{fig4} we plot the SED of the Carina nebula. At 60 and 100 $\mu$m, the fluxes from \emph{COBE} and
from the \emph{IRAS}-IRIS are mutually consistent within 2 and 1$\,\sigma$, respectively. The \emph{IRAS}-IRIS flux at 60$\,\mu$m
is 1$\sigma$ away from \citet{Smith07}, and the COBE one is at 2.9$\sigma$. The \emph{IRAS}-IRIS and the \emph{COBE} fluxes at
100$\,\mu$m are at 2.4 and 3.5$\sigma$, respectively, with respect to \citet{Smith07}. The extrapolation at 3.4$\,$cm from B98
90$\,$GHz and all WMAP7 data is about 10$^3\,$Jy, at 1.6$\sigma$ from the Parkes data at the same wavelength \citep{Smith07}. This
could be due to either an absolute calibration $/$ offset problem in the Parkes data, or the presence of a further background
rapidly decreasing with frequency.

We fitted the SED, excluding the COBE points at 60 and 100$\,\mu$m, using the sum of a graybody (with an emissivity power law
$(\lambda/\lambda_1)^{\alpha_1}$ and amplitude $e_1$) and an optically thin bremsstrahlung component:
\begin{equation}\label{eSed}
F_{\nu}=e_1 \frac{2h\nu^3}{c^2}\frac{1}{(e^{h\nu/kT_1}-1)}(\frac{\lambda}{\lambda_1})^{\alpha_1}\Omega_b+a_s
(\frac{\nu}{\nu_0})^{-0.1},
\end{equation}
where $F_{\nu}$ is the flux density, in Jy, and $\Omega_b$ the area, in sr, of the integration box. A chi-square minimization with
four free parameters ($\chi^2$;4) (the emissivity amplitude at $\lambda_1=100\,\mu$m, $e_1$; the emissivity spectral index
$\alpha_1$; the temperature of the graybody component $T_1$, and the amplitude of the free-free component $a_s$ at the Parkes
frequency $\mathbf{\nu_0}$) gives:
$\chi_{reduced(\chi^2;4)}^2=$1.16, 
$e_{1(\chi^2;4)}=(1.09\pm0.13)\times10^{-3}$, $T_{1(\chi^2;4)}=(35.0\pm2.0)\,$K, $\alpha_{1(\chi^2;4)}=-1.32\pm0.09$, and
$a_{s(\chi^2;4)}=(1100\pm21)$Jy. The bremsstrahlung amplitude is well constrained; the \emph{IRAS}/\emph{COBE} data bind the peak,
while the B98 ones constrain the long wavelength trend of the graybody law.

\begin{figure}
\epsscale{1.00}\plotone{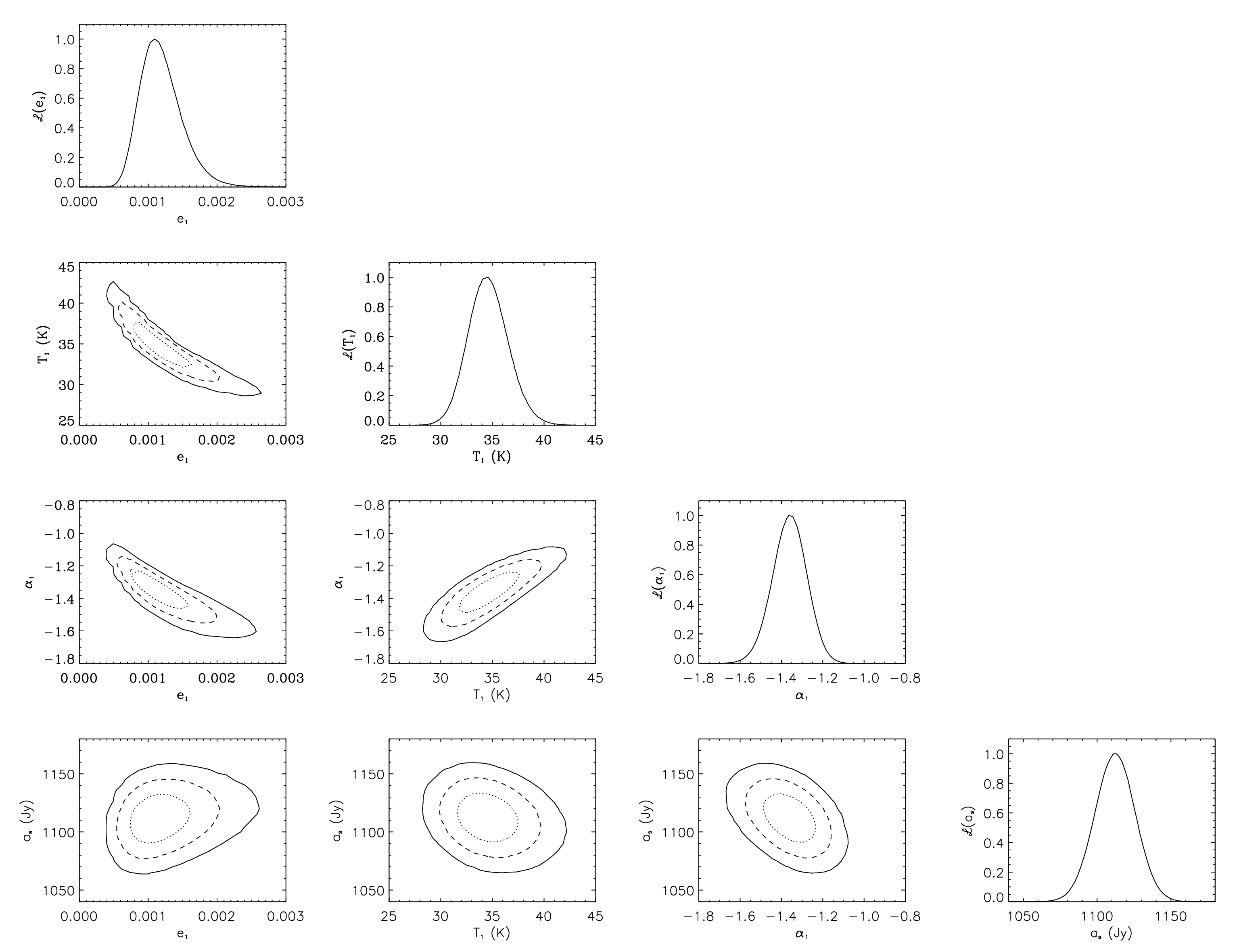} \caption{Contour plots of the joint probability density and likelihoods for the four parameters
describing the SED of the Carina nebula. The dotted, dashed and solid regions represent confidence intervals for 68.3$\%$,
95.4$\%$, and 99.7$\%$ CL, respectively.} \label{fig3}
\end{figure}

\begin{figure}
\epsscale{.80}\plotone{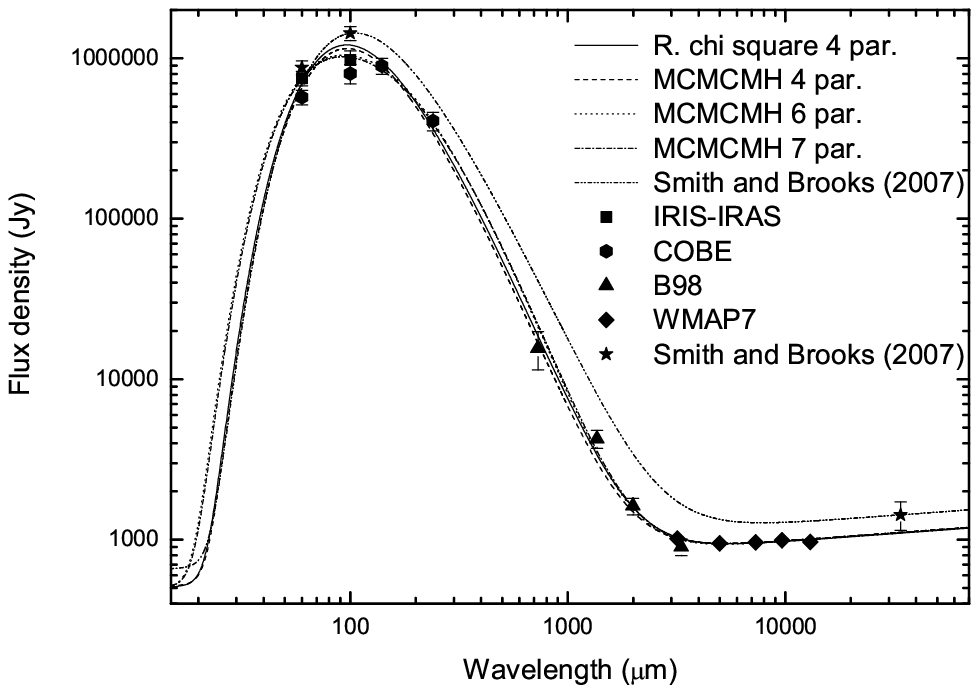} \caption{FIR to radio SED of the Carina nebula. In the legend we report, from top to bottom,: the
best fit SED for four-parameter $\chi^2$ fit, the same for the MCMCMH with four, six, and seven parameters, and the \citet{Smith07}
SED. The symbols represent the measured integrated fluxes.} \label{fig4}
\end{figure}

We studied the correlation between parameters using an MCMCMH algorithm \citep{Andrieu03,Lewis02,MacKay02}. The resulting contour
plots, showing Confidence Levels (CL) regions of 68.3$\%$, 95.4$\,$, and 99.7$\%$ are reported in Figure$\,$\ref{fig3}.

Our parameters are well determined even without the introduction of priors in the algorithm. The graybody parameters are, as
expected, not correlated to the bremsstrahlung one; the temperature, the emissivity amplitude and its spectral index, instead, are
partially correlated. The estimates for the parameters ($mcmc;4$), calculated as the average value of the random numbers accepted
by the algorithm, are: $e_{1(mcmc;4)}=\left( 1.10^{+0.38}_{-0.28} \right) \times 10^{-3}$, $T_{1(mcmc;4)}=\left(34.5^{+2.0}_{-1.8}
\right)\,$K, $\alpha_{1(mcmc;4)}=\left(-1.37^{+0.09}_{-0.08} \right)$, and $a_{s_1(mcmc;4)}=\left( 1113^{+12}_{-15} \right)\,$Jy
with a reduced $\chi^2_{reduced(mcmc;4)}=$1.009. As expected the MCMCMH provides a better chi-square since the algorithm is not
limited by the resolution of the parameters grid step as in a simple grid-based chi-square minimization.

We also investigated the possibility that the SED of the Carina nebula could be described by two graybodies: we added the free
parameters $e_2$, $T_2$, and $\alpha_2$, normalizing the second emissivity power law at $\lambda_2=2000\,\mu$m. The minimization of
the $\chi^2$-square fit provides hints for the possible existence of a colder dust component, with temperature of about 18K and
emissivity of about 2$\times10^{-5}$, while the warmer component would have a temperature of about 40$\,$K, with a lower emissivity
with respect to the case of a single population.

However, the results are unstable because of the large number of free parameters with respect to the number of independent data and
the intrinsic degeneracy between the temperature and the spectral index \citep{Dupac03}; moreover, the improvement of the fit with
respect to the four parameters case is not significant.

We used these results as input for an MCMCMH run with the same seven free parameters. We restricted the parameters space, adding
uniform priors for the conditional probabilities; therefore we set $T_2>0$, $e_2>0$ and $\alpha_2$ varying in the range [-2.4 ;
-0.8]. The latter range has been selected following the temperature dependence of the submillimeter spectral index derived by
PRONAOS data \citep{Dupac03}.

The graybody parameters still remain not correlated to $a_s$, but the small number of experimental data, with respect to the free
parameters, is not sufficient to constrain the parameters of the second graybody and to study their correlation properties. As a
check of the instability of this result, we performed 100 MCMCMH runs, with the same \emph{nstep} and \emph{burn$\_$in}, and
compare the results of the first run to the other 99 ones. Large variations were found, for all but the $a_s$ parameter, and, in
any case, the lowest $\chi^2$ value we found was larger than the previous one.

We tried to eliminate the parameter $\alpha_2$ using its dependence from $T_2$ (see, e.g., \citet{Paradis10, Veneziani10});
following \citet{Dupac03} we tried
\begin{equation}\label{eq DupcaHyp}
\alpha_2=-\frac{1}{0.4+0.008T_2}.
\end{equation}
An MCMCMH with six parameters confirms the results of the MCMCMH with seven parameters of a warmer dust component with a lower
emissivity and a second colder dust population with temperature of about 18$\,$K and emissivity of about 2$\times10^{-5}$. This is
not a significant improvement over the four parameters fit, given the reduction of the degrees of freedom.

We conclude that the data, while not excluding it, do not require an additional cold dust component.

We also checked how the background removal method affects the best fit parameters. In case $n=2$ we find a reduced $\chi^2$ lower
by 10$\%$ with respect to case $n=5$, and a temperature $T_1$ warmer by 2K. All the best fit parameters are consistent with ones
obtained in case $n=5$ within 1$\sigma$.

\bigskip
Integrating the spectral flux density describing the 34.5$\,$K dust population and considering an average distance of the nebula
equal to 2.3$\,$kpc \citep{Allen93}, we calculate the total IR luminosity for this dust population and the corresponding dust mass
(Section$\,$\ref{sec Smith analysis}) obtaining $(7.4^{+2.5}_{-1.4})\times10^6L_{\odot}$ and $(9500^{+4600}_{-3500})M_{\odot}$;
both are well consistent with the corresponding estimates of \citet{Smith07}. In presence of two dust grains populations the total
luminosity is consistent, within 1$\sigma$, with the \citet{Smith07} estimate. On the contrary, the mass determination is strongly
dependent on the assumption of a single dust population. Two dust populations hypothesis does not affect the luminosity because the
SED remains very similar, but has a significant impact on the mass estimation as the mass strongly depends on the temperature. The
nebula mass estimate increases up to 9.2$\times10^5$ solar masses if we add a second population with temperature of about 18$\,$K
(see above).

\section{Discussion and future perspectives}\label{sec discus}

The determination of the SED improves our knowledge of this star forming region. In fact, the spectral index is a signature of a
given type of dust (amorphous carbon and/or porous aggregates of silicates and graphite rather than silicates in ice mantles), and
the parameters describing the SED are used to estimate the luminosity and the mass of a given dust population.

Compared to \citet{Smith07}, our results agree with the presence of a dust population with temperature of 35$\,$K, although the
fits with a double dust population might suggest the presence of slightly warmer dust (about 40$\,$K instead of 35$\,$K). This
results in a significantly lower estimate of the mass for the warm dust component. At variance with \citet{Smith07}, in our
analysis the emissivity spectral index is not assumed a priori, but is obtained from the data. A spectral index around -1.4 was
predicted \citep{Koike80} from laboratory measurements on amorphous carbon, although this can be produced also by silicates,
amorphous carbon and graphite bounded together in a porous aggregate \citep{Mathis89}. A spectral index around -1 was also measured
at 1.3$\,$mm for observations towards two pre-main-sequence stars in the Taurus dark clouds; there thermal emission is due to small
particles embedded in disks of gas and dust surrounding the stars \citep{Beckwith90}. Values between (-1 - 0) and (-2 - -1) are
found, respectively, in submillimeter and millimeter wavelength observations toward T Tauri and FU Orionis stars
\citep{Weintraub89}; the former range is related to extended dust clouds, the latter range is typical of dust in molecular clouds
and compact HII regions. PRONAOS observations of Orion, M17, $\rho$ Ophiuchi and Cygnus B show a spectral index around -1
\citep{Dupac03}, again consistent with our results. Early Planck results for diffuse dust in our Galaxy \citep{Planck11a} show a
spectral index (-1.8), steeper than our value. However, the dust population sampled here is in a different (warmer) environment. In
fact, indications of a steepening of the spectral index for colder temperatures, like in Equation$\,$(\ref{eq DupcaHyp}), have been
noted by several analysis (see e.g. \citet{Paradis10, Veneziani10}).

For our 4 parameters fits, our best fit spectral indexes are: $\alpha_{1(\chi^2;4)}\simeq-1.32$ and
$\alpha_{1(mcmc;4)}\simeq-1.37$. Using the temperatures estimated from the fit, Equation (\ref{eq DupcaHyp}), instead, would
predict -1.47 (1.6$\sigma$ away from ours) and -1.48 ($1.1\sigma$), in general in reasonable agreement with the data.

From a simple equilibrium equation we can estimate the temperature of the grains, $T_g$, as:
\begin{equation}\label{eq equ}
    T_g \sim \sqrt[4]{\frac{<Q_{UV}>}{<Q_{IR}>} \frac{L_{s}}{16\pi \sigma_{SB} d^2}}\sim 40\,K;
\end{equation}
with $<Q_{UV(IR)>}$ the blackbody-averaged Q-factor in the UV (IR) wavelengths \citep{Draine81, Draine84}, $L_s\sim2\times10^7$ the
total stellar luminosity of the Carina nebula \citep{Smith06} and $d\sim40\,$pc (1$^{\circ}$) the distance of the dust grains from
the sources. $<Q_{UV}>$ is the blackbody averaged absorption efficiency (assuming a blackbody temperature of $38\times10^3\,$K);
the typical distance $d$ between the grains and the star has been estimated from the angular size of the nebula and its distance.
This value, although obtained from a naive model, is in reasonable agreement with our results.

Equation$\,$(\ref{eq equ}) can be also used to estimate the mean distance of the dust grains from the center of the nebula once
derived their temperatures. For a dust population of 34.5$\,$K we find a distance of about 50$\,$pc; in the presence of a colder
dust population of about 18$\,$K the distance would be about 200$\,$pc.

\citet{Yonekura05} estimate a mass of the molecular gas of about $3.5\times10^5$M$_{\odot}$; since the molecular gas in the Carina
nebula is about one third of the total mass budget (atomic+ molecular gas+ dust) following \citet{Smith07} we can estimate a total
mass of about 10$^6$M$_{\odot}$; this compared with our estimated mass gives a gas-dust ratio of about 100, which is the canonical
value assumed in this nebula \citep{Morris99, Smith05, Smith07}.

Our best estimate of the free-free emission, based mainly on the \emph{WMAP} fluxes, is at 1.6$\,\sigma$ from the Parkes
observation. One possibility to explain an excess flux observed in the Parkes data would be to suppose the presence of synchrotron
radiation in the nebula, but this does not fit other long wavelength radio observations \citep{Beard66, Jones73}. More likely are
effects related to the very different angular resolution of the Parkes data, 2.6 arcmin \citep{Huchtmeier75}, with respect to the
WMAP observations. With a large beam the flux measured by \emph{WMAP} in the pixels of the box might be overestimated, since every
peripheral pixel includes the contribution of sources outside the integration box, but still inside the fraction of the beam
spilling outside the box. The largest \emph{WMAP} beam, 0$^{\circ}$.81 FWHM, is still smaller than our integration box, so that the
flux of the Carina Nebula lost in the sidelobes of the beam is negligible. However, if the 0$^{\circ}$.8 beam is centered on the
edge of the integration region, it might pick-up flux from sources outside the integration region, where half of the beam solid
angle is lying. We have double checked this statement with numerical simulations. We have simulated a sky brightness distribution
assuming that a single source, 30 arcmin in radius, is located in the center of our integration box. We have convolved this
brightness with Gaussian beams with the same FWHM as the \emph{WMAP} beams and integrated over our integration box to estimate the
measured flux. The result is the same within less than 1$\%$ (at 94 GHz) (or up to 3$\%$ at 23 GHz) as in the case of an unsmoothed
map, showing that the power lost in the sidelobes is negligible. Then we have added a second source, with 30 arcmin diameter and
20$\%$ of the maximum brightness of Carina, just outside our integration box. In this case we observe an overestimate of the
measured fluxes which can be, again, from less than 1$\%$ (at 94 GHz) to 3$\%$ (at 23 GHz) of the unsmoothed map. Therefore, the
modest resolution of the low frequency channels of \emph{WMAP} does not seem to be the reason of the deficit with respect to the
Parkes data. Probably more important is the fact that the Parkes data do not cover our entire integration box \citep{Smith07}. From
all this we conclude that a direct comparison is very difficult. This issue, however, does not affect the dust properties in the
Carina nebula, which are the main focus of this paper.

The Carina nebula SED is well described by a four-parameter fit, although the lack of data between 240 and 730$\,\mu$m cannot
exclude the presence of another dust component. Our 7-parameter fit gives an hint of this; nevertheless more data in the missing
wavelength range are necessary to confirm it.

Our method of averaging over distinct background regions provides a background estimation generally in agreement with what was
found by \citet{Smith07}.

At wavelengths shorter than 60$\mu$m, the emission in the nebula is dominated by polycyclic aromatic hydrocarbons (e.g., Rathborne
et al. 2002) and warm dust grains at temperatures $\gtrsim$80$\,$K. At wavelengths longer than 60$\,\mu$m, the flux from $\eta$
Carinae is negligible with respect to the integrated flux of the nebula. \citet{Cox95} have shown that the composite IR SED of
$\eta$ Carinae and the Homunculus Nebula is described by two modified Planck spectra $\nu^\delta B(\nu,T_j)$ with $\delta=1$ and
$T_j$=210, 430$\,$K. At 20$\,\mu$m, the peak of the flux density from the Homunculus nebula is about 10$^5\,$Jy, while about
10$^2\,$Jy comes from the 35$\,$K dust population of the more extended Carina Nebula \citep{Smith07}; at 60$\,\mu$m the emission of
the Homunculus nebula drops to a value of about 2$\times10^4\,$Jy, while that of the 35$\,$K dust increases to about
9$\times10^5\,$Jy. Among recent results \citep{Smith03}, the ones based on \emph{Infrared Space Observatory} observations
\citep{Morris99} also confirm these findings. The global mass of the Carina nebula is not strongly affected by the mid-IR
($<60\mu$m) portion of the spectrum: the 200$\,$K dust population adds a small contribution to the total mass budget which is
negligible with respect to the large uncertain coming from colder (at 18 or 40$\,$K) dust grains populations.

Our hint for a second, colder, dust population with a temperature of about 18$\,$K would be in good agreement with the analysis of
the star-forming regions observed by PRONAOS, which predicts a high spectral index (about -2) for very cold regions (11-20)$\,$K
\citep{Dupac01,Dupac03}. The analysis of the millimetric continuous emission performed on the SIMBA-SCUBA star-forming regions
shows a typical value of the spectral index equal to -2, assuming dust temperatures higher than 10$\,$K \citep{Hill06}. However,
the existence of a cold component cannot be excluded from the available data. It will be interesting to improve the SED of the
Carina nebula with future observations, both better calibrated and at complementary wavelengths, either confirming or disproving
hints for the existence of the very cold dust component, and its polarization properties.

The future balloon borne experiment PILOT \citep{Bernard07a,Bernard07b} will observe the Carina nebula at the two wavelengths of
240 and 550$\,\mu$m; observations at the first wavelength could confirm the flux measured by COBE-DIRBE, observations at the second
wavelength will fill our missing range.

The ongoing \emph{Planck} mission \citep{Tauber10} has measured the whole sky at nine wavelengths from 30 to 900$\,$GHz, with
unprecedented sensitivity, accurate ($<1\%$) calibration, and with angular resolution down to 5 arcmin; the SPIRE instrument on
board of the Herschel satellite \citep{Griffin10} will also cover the region at 250, 350, and 500$\,\mu$m. Once processed, these
data will certainly solve the issue. The error in the background estimates, at wavelengths shorter than 2$\,$mm, are comparable to
the difference between the SEDs of our one- and two-component best fits and larger than the Planck calibration errors; therefore
they will not be the limiting factor for discriminating between the two models.

\section{Conclusion}\label{sec concl}

We have presented multi-wavelength observations of the Carina nebula. We compared our analysis to the \citet{Smith07} derivation of
the SED. The SED of the nebula can be described as the sum of a graybody component, with a power-law emissivity, and an optically
thin thermal bremsstrahlung. The parameters of the model are well constrained by the data: we find
$e_1=(1.10^{+0.38}_{-0.28})\times10^{-3}$, $T_1=(34.5^{+2.0}_{-1.8})\,$K, $\alpha_1=-1.37^{+0.09}_{-0.08}$,
$a_s=1113^{+12}_{-15}\,$Jy. Thanks to the B98 observations, in the range 730$\,\mu$m - 3.3$\,$mm, we have been able to derive the
spectral index $\alpha_1$ rather than assuming it equal to -1. We also found, from our multi-component analysis, that the data are
consistent with an inverse relation between spectral index and temperature. However, the improvement in the $\chi^2$ is not
statistically significant. Our four-parameter results do not exclude the possibility that the FIR SED is described by two graybody
components, the latter with a colder dust temperature and a steeper spectral index than the former. The small number of
experimental data, compared to the number of free parameters does not allow a robust investigation of this hypothesis. This issue
will be solved by the Planck/Herschel surveys.

\section{Acknowledgements}\label{ack}

This activity has been supported by the Italian Space Agency projects BOOMERanG, OLIMPO, and LSPE, and by University of Rome La
Sapienza.

\end{document}